\documentclass [a4paper,12pt]{article}

\usepackage{amssymb}
\usepackage[centertags]{amsmath}
\usepackage{graphicx}
\usepackage{lscape}
\usepackage{cite}

\newcommand{\br}{\mathbf{r}}

\title{Karhunen-Loeve analysis of complex spatio-temporal dynamics
of thin-films optical system}

\author{M. U. Karelin, P. V. Paulau, I. V. Babushkin \\
 \itshape B. I. Stepanov Institute of Physics, NAS Belarus, \\
 \itshape prosp. F. Skaryny 68, Minsk, 220072 Belarus \\
 \itshape tel: (375 17) 2841419; fax: (375 17) 2840879; \\
 e-mail: karelin@dragon.bas-net.by}

\date{}

\begin{document}

\maketitle

\begin{abstract}
Application of Karhunen-Loeve decomposition (KLD, or singular
value decomposition) is presented for analysis of the
spatio-temporal dynamics of wide-aperture vertical cavity surface
emitting laser (VCSEL), considered as a thin-layer system. KLD
technique enables to extract a set of dominant components from
complex dynamics of system under study and separate them from
noise and inessential underlying dynamical behavior. Properties of
KLD spectrum and structure of its main components are studied for
different regimes of VCSEL. Along with the analysis of VCSEL, a
brief survey of KLD method and its usage for theoretical and
experimental description of nonlinear dynamical systems is
presented.
\end{abstract}

\\[5mm]
% from Loiko & Babushkin article
\noindent\textbf{UDC:} 621.373.826.038+539.2

%--------------------------------------------------------------------
\section{Introduction}

Investigation of interaction of thin-film systems with laser
radiation becomes quite topical during the last decade. This is
mainly stimulated by advances of semiconductor technology, which
enable to obtain multi-layer semiconductor structures with
thickness close to or even less than wavelength of visible
radiation. In particular, wide-aperture vertical cavity surface
emitting lasers (VCSELs) are wide-spread optical systems of
communication and information processing. Complex spatio-temporal
regimes of VSCEL operation, in principle, opens new possibilities
for information processing (for example, ``chaotic'' encoding for
secure applications). On the other hand, it is far from full
understanding, what mechanisms form spatial and temporal structure
of radiation in VCSELs.

In the paper, we investigate the spatiotemporal dynamics of VCSEL
using the system of differential equations describing the dynamics
of broad area VCSEL \cite{Loiko2001}. The equations are derived
using spin flip model \cite{SanMiguel95, MartinRegalado97} and
take into account polarization of light and complex cavity of
VCSEL, including Bragg reflectors. %(see Section~\ref{model}).
The propagation of radiation through the VCSEL cavity is
calculated in an approximation of thin nonlinear layer, which
allows to dismiss the diffraction of light in the active medium.
Calculations of the current intensity $J$ near threshold and for
some values far from threshold are performed for the case of zero
phase anisotropy, low amplitude anisotropy and homogeneous spatial
profiles of refraction index and injection current.

Obtained sets of spatiotemporal data are analyzed using
Karhunen-Lo\'eve decomposition (KLD, also known by several other
names \cite{Holmes_etal}). Such technique is introduced in the
beginning of 20th century for description of random functions.
Then this method have found numerous applications in such areas as
pattern recognition, turbulence, meteorology, coherence theory
etc. Optimal properties of KLD enables to extract only a few main
components from the whole complex dynamics of system under study.

The rest of article is organized as follows: in the next section
we outline the method of or Karhunen-Lo\'eve decomposition and its
main characteristics. In the third section we provide the
mathematical model of VCSEL is presented. In the following section
we analyze of complex dynamics of VCSEL by KLD method. In
particular, change of the decomposition spectrum for the values of
the current density $J$ near lasing threshold and for some values
above threshold is discussed. The last section contains a
conclusion and outline of future tasks.

%--------------------------------------------------------------------
\section{Karhunen-Lo\'eve decomposition}

%%%%%%%%

In its simplest form, suitable for our purposes, the KLD method is
formulated as follows: given some (in general, complex) field of
two variables $u(\br,t)$ with $\br = (x,y)$, one tries to find its
decomposition onto purely temporal and spatial modes:
 \begin{equation}
 \label{eq-KLD}
 u(r,t) = \sum_{i=1}^{\infty} \lambda_i \, a_i(t) \, \phi_i(\br).
 \end{equation}
with two orthonormality conditions
 \begin{equation}
 \label{eq-KL-ort}
 \int_S d\br \; \phi_i(\br) \, \phi^{\ast}_j(\br) = \delta_{ij}
 \qquad
 \int_T dt \; a_i(t) \, a^{\ast}_j(t) = \delta_{ij}
 \end{equation}
where $T$ is a time interval and $S$ is an area  on which we want
to analyze field. This field is considered as known, e.g. from
experiment, or from some kind of model - either analytical or
numerical.

The physical sense of representation (\ref{eq-KLD}) is extraction
of spatial distributions which oscillate in time as a whole. The
values $|\lambda_i|^2$ give the part of `energy' carried by $i$-th
mode in average. Physical origin of $u(\br,t)$ is of little
importance --- it could be electrical field or intensity of some
kind of radiation, velocity profile of flow or even simply set of
pictures $u(\br)$ numbered by second variable $t$
\cite{EversonSirovich, Kurashov}. In our case, $u(\br)$ is the
slowly varied complex envelope of the optical field. One of the
main parameters of presentation (\ref{eq-KLD}) is the number of
terms $N_\epsilon$, such that their sum contains not less that
some prescribed part of the whole energy of $u(\br,t)$
 \begin{equation}
 \label{eq-KL-N}
 \sum_{i=1}^{N_\epsilon} |\lambda_i|^2 \geqslant
 (1-\epsilon) \sum_{i=1}^{\infty} |\lambda_i|^2.
 \end{equation}
The other components with $i > N_\epsilon$ bears noise or
unimportant dynamics and so could be excluded from consideration
(on given interval of time and spatial domain).

As it could be shown, the decomposition functions could be found
from two eigenproblems for integral equations
 \begin{equation}
 \label{eq-KL-spat}
 |\lambda_i|^2 \, \psi_i(\br) = \int_S d\br' \; \phi_i(\br') \, K_S(\br,\br')
 \end{equation}
and
 \begin{equation}
 \label{eq-KL-temp}
 |\lambda_i|^2 \, a_i(t) = \int_T dt' \; a_i(t') \, K_t(t,t')
 \end{equation}
where kernels are correlation function
 \begin{equation}
 \label{eq-kern-spat}
 K_S(\br,\br') = \int_T dt \; u(\br,t) \, u(\br',t)
 \end{equation}
and some kind of temporal correlation function, averaged over
space
 \begin{equation}
 \label{eq-kern-temp}
 K_S(\br,\br') = \int_T dt \; u(\br,t) \, u(\br',t).
 \end{equation}
Equation (\ref{eq-KL-temp}) with kernel (\ref{eq-kern-temp}) is
usually referred to as ``method of snapshots'' of ``method of
strobes'' \cite{Holmes_etal}.

On the other hand, decomposition of type (\ref{eq-KLD}),
(\ref{eq-KL-ort}) corresponds to singular-value decomposition
\cite{SIAM} of ``matrix'' $u(\br,t)$. As far as experimental (or
numerical calculation) data is always is a kind of matrix of
numbers. Appropriate discrete decomposition may be effectively
calculated using standard \texttt{svd} routine available in umber
of mathematical packages.

It should be noted, that whole decomposition could be found from
just one eigenproblem, while the dual basis, temporal or spatial,
is found from projection
 \begin{equation}
% no label
 \notag
 \lambda_i \, a_i(t) = \int_S d\br \; \phi^{\ast}_i(\br) \, u(\br,t),
 \end{equation}
 \begin{equation}
% no label
 \notag
 \lambda_i \, \psi_i(\br) = \int_T dt \; a^{\ast}_i(t) \, u(\br,t).
 \end{equation}

Use of eigenfunctions related to the investigated field cause main
positive sides of Karhunen-Lo\'eve expansion: among all sums of
type (\ref{eq-KLD}) with finite number of terms, the
representation in terms of eigenfunctions ensures minimal least
square error and the maximal capture of "energy." The KLD method
proves its power on number of problems in different areas of
physics and other sciences. On the other hand, its main drawback
again related with use of eigenfunctions: analytical solution of
equations (\ref{eq-KL-spat}), (\ref{eq-KL-temp}) is known only for
very few special cases, the numerical solution often require too
much resources and is incapable to provide all the information
about
system dynamics, especially near critical points. To this point, % ???
it is important to look for methods of analysis, which provide
information about spectrum of eigenvalues $|\lambda|^2$ without
calculation of decomposition itself \cite{LazarukKarelin}.

However, calculation of Karhunen-Lo\'eve decomposition provides
valuable information about the details of complex process.
Solution of eigenvalue problem is usually much more easy task than
to study other, ``standard'' parameters of chaotic systems, such
as Lyapunov exponents or fractal dimensions. In most cases
singular-value analysis enables to select just the few most
important components from whole spatio-temporal dynamics and to
study their behaviour.

%--------------------------------------------------------------------
%\section{Two thin-film system and its analysis}

% omit in this preprint

%--------------------------------------------------------------------
\section{Short description of the VCSEL model}
\label{model}

Response of the active medium to the radiation in  VCSEL is in a
semiclassical approximation is described by the spin-flip model
\cite{SanMiguel95, MartinRegalado97, Loiko2001}, taking into
account vector character of field:

 \begin{equation}
 \label{spinflip}
 \left\{
   \begin{aligned}
     \frac{d \mathbf{P}}{dt} & = - \left(\frac{1}{T_2}+i \delta \right)
     P - \frac{|d|^{2}}{3 \overline{h}} i A \mathbf{E},
   \\
     \frac{N}{dt} & = -\frac{N-J}{T_1} - \frac{i}{2 \overline{h}}
     \left(\mathbf{E^{*}} \mathbf{P} - \mathbf{E} \mathbf{P^{*}} \right),
   \\
     \frac{n}{dt} & = - \gamma_s n + \frac{1}{2 \overline{h}}
     \left(\mathbf{E^{*}} \mathbf{P'} - \mathbf{E} \mathbf{P'^{*}} \right).
   \end{aligned}
 \right.
 \end{equation}
where $\mathbf{P}$ is the polarization of the two level centers,
$\mathbf{P'}$ is the vector with components $(P_{y},-P_{x})$; $N$
is the total population difference between the conduction and the
valence bands and $N_{0}$ is its transparency value; $n$ is the
difference between the population differences for the two allowed
transitions between magnetic sublevels, $\delta = \omega_g -
\omega_c$ is the detuning  between the bandgap frequency
$\omega_g$ and the cavity resonance $\omega_c$; $\gamma_s$ is the
decay rate between the magnetic sublevels; $T_1$ and $T_2$ are the
relaxation times for the total population difference and the
polarization correspondingly; $|d|$ is the absolute value of the
dipole momentum of the transition (we suppose it is the same for
both transitions); $J$ is the pump parameter, and
 \begin{equation}
 \label{AMatrix}
 A = \left (
 \begin{array}{cc}
 N-N_{0} & in \\
     -in & N-N_{0}
 \end{array}
 \right ).
\end{equation}

We will use equations (\ref{spinflip}) with adiabatically
eliminated polarization. The procedure of such adiabatic
elimination is described in details in \cite{Loiko2001} and allows
to take into account the asymmetry of the gain line using so
called linewidth enhancement factor $\alpha$. In addition, this
procedure allows to avoid the short-frequency instabilities
intrinsic to the straightforward adiabatic elimination procedure
for the spatially extended lasers. In this approximation the
polarization of the active medium is defined as follows:
 \begin{equation}
 \label{ElimPolariz}%
 P = -\frac{|d|^{2} T_{2}}{3 \overline{h}} (i-\alpha) D
 \hat{\pounds} \mathbf {E},
 \end{equation}
here $D=N-N_{0}$, the operator $\hat{\pounds} \sim
\pounds(\vec{k}_{\perp})= 1/[1 + T_2^2 (\delta -
\Omega(\vec{k}_{\perp}))^2]$, where $\Omega (\vec{k}_{\perp})$ is
a cavity resonance frequency for the tilted wave with definite
$\vec{k_{\perp}}$, and the tilde means an equivalence in the sense
of transverse Fourier transform.

The propagation of radiation through the VCSEL cavity is
calculated in an approximation of thin film active medium, which
allows to neglect the diffraction in an active layer
\cite{Loiko2001}. It gives us the following relation:
 \begin{equation}
 \label{Ecorrelation}
 \mathbf {E_{i}} = \hat{F} \: \mathbf {E_{t}}.
 \end{equation}
where $\mathbf{E_i}$ is the field incident into the active medium,
$\mathbf{E_t}$ is the field outgoing from the active medium.
Operator $\hat{F} = \rho \, \exp({2ikL + i (\Delta_{\perp}/k) L})
\, \Gamma \, \hat{R}$, where $\rho$ describes absorption in the
linear medium between the active layer and reflector, $L$ is the
thickness of the spacer layer, $\Delta_{\perp}$ is the Laplasian
in the transverse plane,
 \begin{equation}
 \label{GMatrix}
 \Gamma = \left (
   \begin{array}{cc}
     e^{\gamma_{a}+i \gamma_{p}} & 0  \\ 0 & e^{-(\gamma_{a}+i
     \gamma_{p})}
   \end{array}
 \right ),
 \end{equation}
is an polarization anisotropy matrix, with $\gamma_a$, $\gamma_p$
being the amplitude and phase anisotropy parameters
correspondingly. Operator $\hat{R}$ describes the reflection of
plane waves from the Brag reflectors \cite{Loiko2001}.

For numerical simulations we used the following parameters:
$\gamma_{s}=100$, $\gamma_{a}=0.1$, $\gamma_{p}=0.0$, $\delta=
0.006$, $\alpha = 3.0$. For chosen parameters  the lasing
threshold is $J_{0}=0.730$ which is needed for the following
discussion (see \cite{Loiko2001} for the detailed description of
threshold conditions).

%--------------------------------------------------------------------
\section{Analysis of VCSEL dynamics}

%\subsection{The regular spatiotemporal regime near threshold}
%\label{sec:regular}

\begin{figure}[!h]
\begin{center}
\includegraphics[width=10cm]{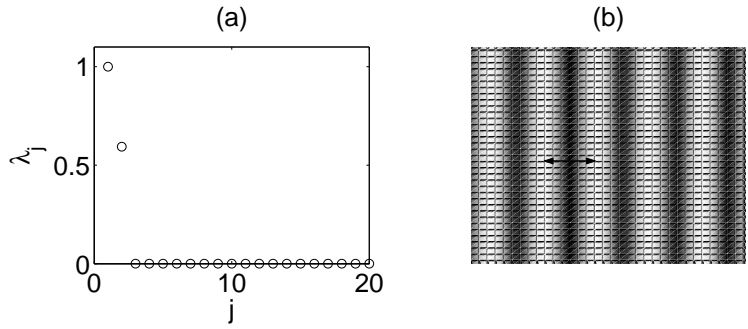}%%
\caption{(a) - The KL spectra for J=0.740. (b) - The zoomed
snapshoot of the dynamics. The arrows shows the direction of
oscillations(the amplitude of oscillations is smaller then the
length of arrows).} \label{klnearth}
\end{center}
\end{figure}

We consider first the simplest homogeneous case with periodic
boundary conditions and with small amplitude anisotropy. The
resulting spatiotemporal regime near threshold ($J=0.740$) is
regular and consists of stripes that weakly oscillate near certain
equilibrium state (the directions of oscillations is shown in
Fig.\ref{klnearth}b by arrows). The contrast of stripes also
changes during the evolution. The KL-spectra of eigenvalues
$\lambda_j$ of the spatiotemporal regime for this case is
presented in Fig.\ref{klnearth}a. It is evident, that only two
modes are the most significant, while all the others could be
safely treated as zeros.

\begin{figure}[!h]
\begin{center}
\includegraphics[width=10cm]{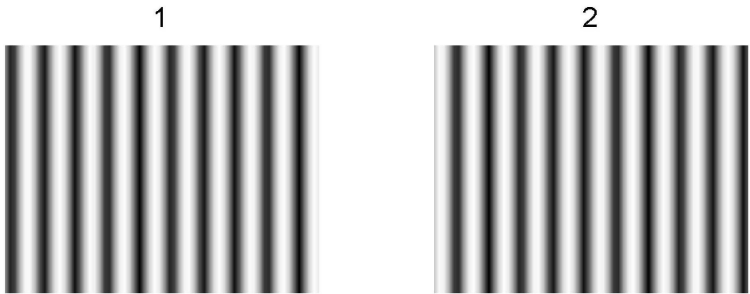}%%
\caption{ The first 2 spatial KL-modes for $J=0.740$.}
\label{sklnearth}
\end{center}
\end{figure}

\begin{figure}[!h]
\begin{center}
\includegraphics[width=10cm]{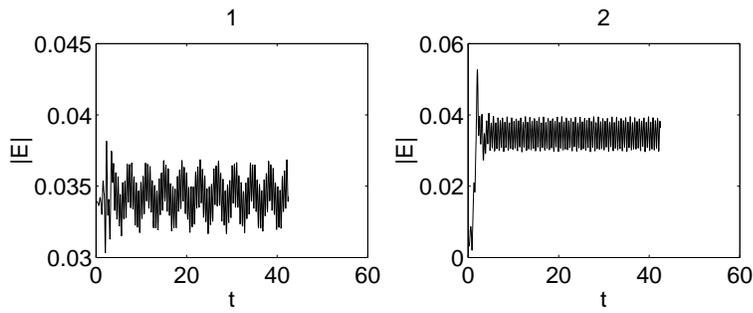}%%
\caption{ The temporal evolution of KL-modes shown in Fig.
\ref{sklnearth}.} \label{tklnearth}
\end{center}
\end{figure}

The spatial KL-modes appear to be stripes (Fig.\ref{sklnearth}).
Their maxima don't coincide and thus a spatial phase shift leads
to orthogonality and to spatiotemporal dynamics. Hence, the
averaging over a large period of time leads to stripes with
smaller contrast. Time dependencies of the KL-modes (see Fig.
\ref{tklnearth}) also characterized by regular (oscillating)
dynamics, except only some transient stage.

\begin{figure}[!ht]
\begin{center}
\includegraphics[width=10cm]{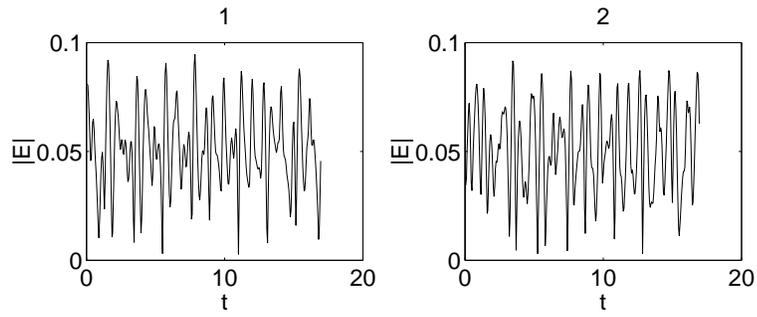}%%
\caption{ The first 2 temporal KL-modes for J=0.800.}
\label{tkl800}
\end{center}
\end{figure}

However, a slight enlarge of the injection current (up to
$J=0.800$) leads to chaotic time dependency the field.
Nevertheless, in this case the KL-spectra has the same form as one
near threshold Fig.~\ref{klnearth}a, and, moreover, the spatial
KL-modes are the same as in Fig.~\ref{sklnearth} too. However,
their temporal behavior is sufficiently different (see
Fig.\ref{tkl800}).

Therefore, the chaotization of the regular dynamics appears due to
mechanism, which is not connected with excitation of the
long-wavelength inhomogenities, either Eckhaus or zig-zag type
(i.e. in $x$ or $y$ directions).

\begin{figure}[!ht]
\begin{center}
\includegraphics[width=7cm]{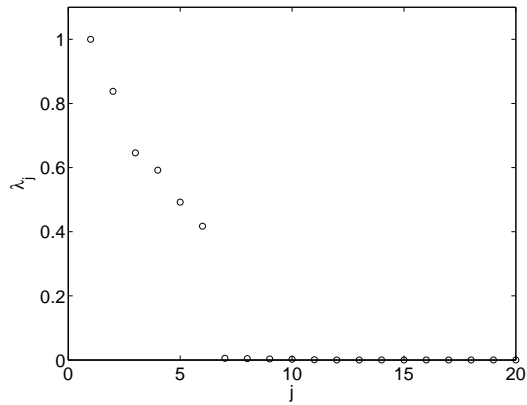}%% j1.000homog_klspectr.eps
\caption{The KL spectra for J=1.000.} \label{klsfar}
\end{center}
\end{figure}

\begin{figure}[!ht]
\begin{center}
\includegraphics[width=10cm]{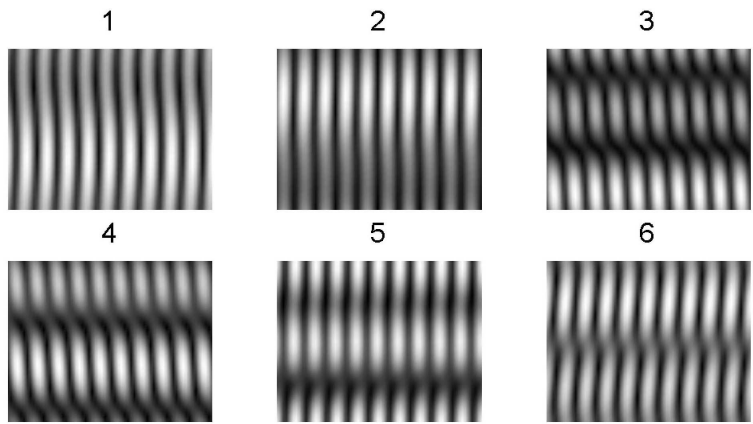}%% j1.000homog_smodes.eps
\caption{ The first 6 spatial KL-modes for J=1.000.}
\label{klsmodesfar}
\end{center}
\end{figure}

Further increase of the injection current ($J=1.000$) leads to
more disordered dynamics. More than two active modes are present
in in KL-spectrum (Fig.~\ref{klsfar}). The spatial structure is
also changing in this case, and modulation in $x$ direction now is
accompanied by modulation in $y$ direction (see
Fig.~\ref{klsmodesfar}).

%--------------------------------------------------------------------
\section{Conclusion}

In summary, our calculations have shown, that during transition of
VCSEL's from the regular behavior to spatio-temporal chaos, it
still can be described by superposition of just a few modes with
relatively simple structure (for moderate values of the injection
current). Increase of order parameter (injection current $J$)
leads to activation of some new modes, with new features of
transversal and temporal dependence.

Hence, the observed chaos is not truly ``spatio-temporal''.
Complex dynamics in time domain (both for the electomagnetic field
itself and for KL-modes) is accompanied by just very simple
spatial structure of modes. Moreover, the whole dynamics is
described by only a few components. This fact, together with
importance of VCSELs in modern optical communication, enables to
suppose, that chaotic regimes could be effectively controlled by
adjusting parameters of a system.

%However, taking into account rather long time necessary to model
%the dynamics, it would be preferable to continue the research
%using slightly simplified system of equations together with more
%sophisticated calculation technique (in the present paper we
%utilized Fourier method).

%--------------------------------------------------------------------
\section*{Acknowledgement}

This research was partially supported by Deutsche
Forschungsgemeinschaft (DFG --- German Research Foundation) under
project 436 WER 113/17/2-1. Authors would also thank Dr. Thorsten
Ackemann and Dr. Natalia Loiko for stimulating discussions.

%--------------------------------------------------------------------

%--------------------------------------------------------------------
\newpage

\end{document}